\documentclass[]{emulateapj}

\newcommand {\omegapez}	{\omega_\mathrm{pe,0}}

\shorttitle{COLLISIONLESS SHOCKS IN UNMAGNETIZED PAIR PLASMAS}
\shortauthors{KATO}

\begin{document}

\title{Relativistic Collisionless Shocks in Unmagnetized Electron-Positron Plasmas}

\author{Tsunehiko N. Kato\altaffilmark{1}}
\affil{National Astronomical Observatory of Japan,
2-21-1 Osawa, Mitaka, Tokyo 181-8588, Japan}
\altaffiltext{1}{
Present address:
Institute of Laser Engineering, Osaka University,
Yamada-oka 2-6, Suita, Osaka 565-0871, Japan
}
\email{kato-t@ile.osaka-u.ac.jp}

\begin{abstract}
It is shown that collisionless shock waves can be driven
in unmagnetized electron-positron plasmas
by performing a two-dimensional particle-in-cell simulation.
At the shock transition region,
strong magnetic fields are generated by a Weibel-like instability.
The generated magnetic fields are strong enough to deflect
the incoming particles from upstream of the shock at a large angle
and provide an effective dissipation mechanism for the shock.
The structure of the collisionless shock propagates at an almost constant speed.
There is no linear wave corresponding to the shock wave and therefore
this can be regarded as a kind of ``instability-driven'' shock wave.
The generated magnetic fields rapidly decay in the downstream region.
It is also observed that 
a fraction of the thermalized particles
in the downstream region
return upstream through the shock transition region.
These particles interact with the upstream incoming particles and
cause the generation of charge-separated current filaments
in the upstream of the shock as well as
the electrostatic beam instability.
As a result, electric and magnetic fields are generated
even upstream of the shock transition region.
No efficient acceleration processes of particles were observed in our simulation.
\end{abstract}


\section{Introduction}
Collisionless shocks are driven
in various situations in astrophysical plasmas.
They often accelerate particles and generate
nonthermal high-energy particles.
Such shocks in supernova remnants are known
as the accelerators of nonthermal electrons
\citep[e.g., SN1006;][]{Koyama95},
and are also believed to be the accelerators of cosmic rays
with energies up to the knee energy ($\sim 10^{15.5}$ eV).
Collisionless shocks also exist
in electron-positron plasmas,
for example, that in the Crab nebula,
and they are considered to generate nonthermal,
high-energy electrons and positrons as well.

Recently,
it has been suggested from observations that
the magnetic fields are amplified or generated
around collisionless shocks in several supernova
remnants \citep{Vink03, Bamba03, Voelk05}.
These magnetic fields may be generated by
the high-energy particles accelerated at the shocks \citep[e.g.,][]{Bell04}.
On the other hand,
such a generation mechanism
can be related to the microphysics of the collisionless shocks themselves.
For the generation of magnetic fields
in the relativistic shocks associated with
the afterglows of gamma-ray bursts,
\citet{Medvedev99} suggested that
the Weibel instability \citep{Weibel59, Fried59}
is driven at the shock and generates
strong magnetic fields.
This mechanism can work in shocks
in electron-positron plasmas as well.

For electron-positron plasmas,
the Weibel instability has been investigated in detail
by using numerical simulations.
Several particle-in-cell simulations showed that
in counterstreaming electron-positron plasmas,
the Weibel instability develops
and generates a strong, sub-equipartition magnetic field
\citep{Kazimura98, Silva03}.
Furthermore,
some authors found that
in their simulations with longer calculation times,
a shock-like structure that associates with a strong magnetic field
is formed
even in unmagnetized electron-positron plasmas
\citep{Gruzinov01, Haruki03}.
Thus,
it should be clarified whether
the shock-like structure can be regarded as a shock
from a macroscopic point of view,
for example,
whether the structure dissipates the upstream bulk kinetic energy
at the ``shock'' transition region
or propagates at a steady speed
into the upstream plasma, etc.,
as well as other detailed features.
If they are considered as shocks,
it is also important to elucidate
the role of the strong magnetic fields
generated by the Weibel-like instability
in the dissipation of the shock.

In this study,
we further investigate the dynamics of collisionless shocks
in unmagnetized electron-positron plasmas in detail
by performing a high-resolution two-dimensional particle-in-cell simulation
and confirm that a kind of collisionless shocks indeed forms
mainly due to the magnetic fields generated by the Weibel-like instability.
However, we also find
that no efficient particle acceleration occurs in the shock
and the generated magnetic fields rapidly decay within the shock transition region,
while in the real world, particles are accelerated at the shocks and
large-scale, sub-equipartition magnetic fields are generated
around the shocks.
The reasons why these processes are not observed in our simulation may be
related to the background magnetic field, the spatial and temporal scales
of the simulation, or the composition of the plasma.

\section{Simulation}
In order to investigate the collisionless shocks
in electron-positron plasmas without background magnetic fields,
we performed numerical simulations.
The simulation code is a relativistic, electromagnetic, particle-in-cell code
with two spatial and three velocity dimensions (2D3V)
developed based on a standard method described by \cite{Birdsall}.
Thus,
the basic equations of the simulation are
the Maxwell's equations (in Gaussian units):
\begin{equation}
\frac{1}{c} \frac{\partial \mathbf{E}}{\partial t} = \nabla \times \mathbf{B} - \frac{4\pi}{c} \mathbf{J},
\qquad
\frac{1}{c} \frac{\partial \mathbf{B}}{\partial t} = -\nabla \times \mathbf{E},
\end{equation}
with
\begin{equation}
\nabla \cdot \mathbf{E} = 4\pi \rho,
\qquad
\nabla \cdot \mathbf{B} = 0,
\end{equation}
as constraints,
where $c$ is the speed of light;
$\mathbf{E}$ and $\mathbf{B}$, the electric and magnetic fields, respectively;
$\mathbf{J}$, the current density;
and $\rho$, the charge density,
together with the equation of motion of particles:
\begin{equation}
\frac{d \mathbf{u}}{dt}
=
\frac{q}{mc} \left( \mathbf{E} + \frac{\mathbf{u} \times \mathbf{B}}{\gamma} \right),
\end{equation}
where $q$ and $m$ are the charge and mass of the particle, respectively;
$\gamma$, the Lorentz factor of the particle;
and $\mathbf{u} \equiv \gamma \mathbf{v}/c$, the 4-velocity of the particle
(where $\mathbf{v}$ is the ordinary 3-velocity).
Particle velocities are often denoted in terms of the 4-velocity
throughout this paper.

In the following,
we consider $\tau = \omegapez^{-1}$ as the unit of time
and the electron skin depth $l_0 = c \omegapez^{-1}$ as the unit of length,
where $\omegapez \equiv (4\pi n_{e0} e^2 / m_e)^{1/2}$
is the electron plasma frequency defined for the mean electron
number density $n_{e0}$;
$m_e$ is the electron mass,
and $-e$ is the electron charge.
The units of electric and magnetic fields are both taken as
$E_* = B_* = c (4\pi n_{e0} m_e)^{1/2}$;
they are defined so that
their corresponding energy densities
are both equivalent to the half of
the mean electron rest mass energy density.

The simulations were performed
on a grid of $N_x \times N_y = 4096 \times 512$
with $3.2 \times 10^8$ particles for each species.
The physical size of the simulation box is $L_x \times L_y = 480 \times 60$
in the unit of $l_0$.
Thus, the size of a cell is $(\Delta x, \Delta y) = (0.12, 0.12)$ in the same unit.
The time step is taken as $\Delta t = 0.025$.

Since we consider collisionless shocks in
unmagnetized plasmas,
the electromagnetic fields were initially set to zero
over the entire simulation box.
The boundary condition of the electromagnetic field
is periodic in each direction.

In this simulation,
a collisionless shock wave is driven
according to the well-known ``injection method.''
There are two walls at $x=30$ and $450$.
They reflect particles specularly,
but let the electromagnetic waves pass through freely;
the electromagnetic waves propagating
outside the walls ($0 < x < 30$ and $450 < x < 480$)
are dissipated by means of Ohmic dissipation
so that they cannot reach the wall at the other side.
Initially, both the electrons and positrons are loaded uniformly
in the region between the two walls
with a bulk velocity of $u_x = 2.0$
(thus, the corresponding bulk Lorentz factor is $\Gamma = 2.24$).
Their thermal velocities were given so that
the distribution of each component of the 4-velocity $u_i$ ($i = x, y, z$)
obeys the Gaussian distribution
with a standard deviation of $\sigma_\mathrm{th} = 0.1$ in the plasma
 rest frame, or the ``fluid'' frame.
(In effect, this generates the Maxwell distribution.)
At the initial stage of the simulation,
particles that were located near the right wall (at $x = 450$)
were reflected by the wall
and then interacted with the incoming particles,
i.e., the upstream plasma.
This interaction causes some instabilities
and then a collisionless shock is formed.
It should be noted that the frame of the simulation
corresponds to the downstream rest frame.
Thus,
we observed the propagation of the shock
from right to the left in the downstream rest frame.

\section{Results}
The following figures show the quantities at the end of the simulation ($t = 233$)
otherwise stated.

\subsection{Basic Features of Collisionless Shock}
\begin{figure}
\plotone{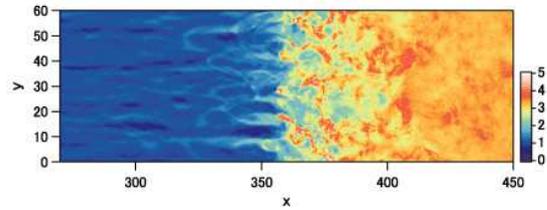}
\caption{ 
\label{fig:ne}
		Electron number density at $t=233$.
		The value is normalized to the far upstream number density
		measured in the downstream rest frame (i.e., the simulation frame).
		The left-hand side and the right-hand side of this figure correspond to
		upstream and downstream of the shock, respectively.
		The transition region extends from $x=350$ to $400$.
}
\end{figure}

Figure~\ref{fig:ne} shows the electron number density 
normalized to the initial number density,
or the far upstream number density,
measured in the downstream rest frame, $n_{e0}$.
In this figure,
the left-hand side and the right-hand side
correspond to upstream and downstream
of the shock, respectively.
We observe that the number density rapidly increases
while crossing the transition region, which extends roughly from $x=350$ to $400$,
and then it becomes homogeneous by moving further downstream.
The transition region does not have
a one-dimensional structure
but a two-dimensional filamentary  structure with
density fluctuations along the $y$ direction.
The typical size of the filaments
is in the order of several electron skin depths.

\begin{figure}
\plotone{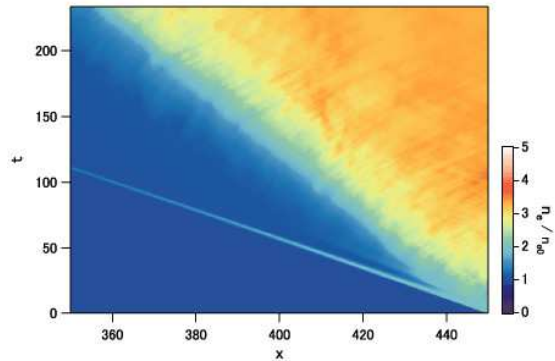}
\caption{ 
		\label{fig:ne_history}
		Time development of the electron number density.
		The horizontal axis is the $x$ coordinate
		and the vertical axis is the time.
		The plotted electron number density is
		averaged over the $y$ direction.
		[\textit{See the electronic edition of the Journal for a
		color version of this figure}.]
}
\end{figure}

Figure~\ref{fig:ne_history} shows the time development of the electron number density.
The horizontal and vertical axes represent the $x$ coordinate and time, respectively.
The plotted number density is averaged over the $y$ direction.
We observe that
the transition region, which is visible as the jump in the number density,
propagates upstream, i.e., to the left, with time at an almost constant speed.
From this figure,
the propagation speed measured in the downstream rest frame
is estimated to be $V_\mathrm{sh,d} \sim -0.39c$.

The thin structure propagating upstream at almost the speed of light,
which is visible in the lower-left portion of this figure,
is due to the particles reflected at the right wall
at a very early stage of the simulation.
This is of course the consequence of the initial condition,
but such a situation may be realized in real cases.
Anyway,
this structure fades away with time,
and it would not affect the structure around the transition region later.

\begin{figure}
\plotone{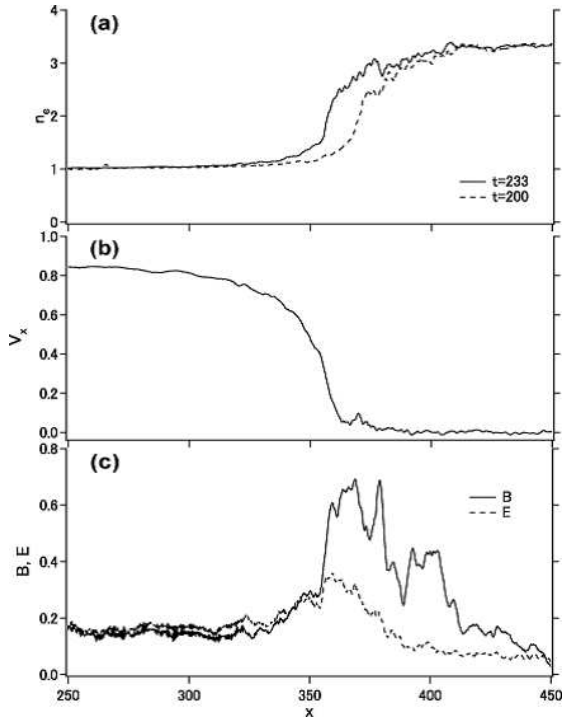}
\caption{ 
		\label{fig:profile}
		The $y$-averaged profiles of quantities at $t=233$:
		(a) the electron number density $n_e/n_{e0}$
		(solid curve for $t=233$ and dashed curve for $t=200$),
		(b) the mean velocity in the $x$-direction $V_x/c$,
		and (c) the magnetic field $|B|/B_*$ (solid curve)
		and electric field $|E|/E_*$ (dashed curve).
}
\end{figure}

Figure~\ref{fig:profile} shows the $y$-averaged profiles of three quantities
around the shock transition region:
(a) the electron number density, $n_e / n_{e0}$;
(b) the mean velocity of electrons in the $x$-direction, $V_x / c$;
and (c) the magnitude of electric and magnetic fields, $|E| / E_*$
and $|B| / B_*$.
In Fig.~\ref{fig:profile}a,
the solid curve and the dashed curve represent the number densities
at $t = 233$ and $t=200$, respectively.
Comparing the two curves,
we see that the transition region propagates  upstream keeping its averaged structure almost unchanged.
The compression ratio measured in the downstream rest frame is approximately $3.3$.
The mean velocity of electrons in the $x$ direction, $V_x$,
rapidly approaches zero within the transition region (Fig.~\ref{fig:profile}b).
In Fig.~\ref{fig:profile}c,
it is notable that a strong magnetic field exists
within the transition region ($350 < x < 400$).
We also see that
behind the shock transition region,
the strength of the magnetic field decreases
and strong electric and magnetic fields exist
even in the upstream region.
These points will be mentioned later.

\begin{figure}
\plotone{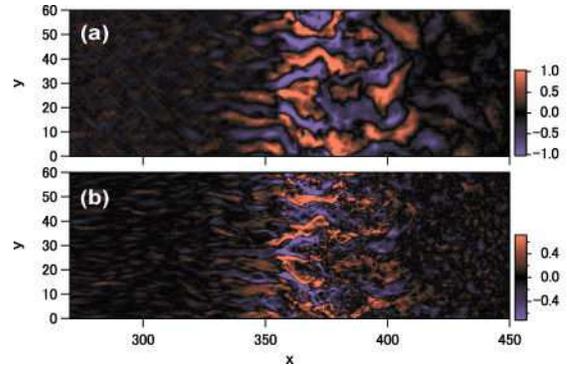}
\caption{ 
		\label{fig:Bz_Jx}
		(a) The $z$ component of the magnetic fields $B_z$ and
		(b) the $x$ component of the current density $J_x$ at $t=233$.
}
\end{figure}

Figure~\ref{fig:Bz_Jx}a shows
the $z$ component of the magnetic field, $B_z$.
It is clear that a strong magnetic field exists
within the transition region ($350 < x <400$),
as indicated in Fig.~\ref{fig:profile}c.
Figure~\ref{fig:Bz_Jx}b shows
the current density in the $x$ direction, $J_x$.
We observed that
there are considerable number of current filaments that carry currents in the $x$ direction
within the transition region.
The filamentary structure observed in the number density (Fig.~\ref{fig:ne})
is in fact due to the existence of these current filaments.
They generate a magnetic field fluctuating in the $y$ direction,
as shown in Fig.~\ref{fig:Bz_Jx}a.
This situation
in which strong magnetic fields fluctuating perpendicular to the direction
of streaming are generated is similar to those in the simulations
of the Weibel instability in counterstreaming plasmas
\citep{Kazimura98, Silva03}.
Therefore,
it is reasonable to consider that
in the transition region
in which the upstream plasma is mixed up with the downstream plasma,
a large velocity dispersion in the $x$ direction and hence
a large anisotropy in the particle velocity distribution are induced;
further, the anisotropy drives
the Weibel-type instability, which generates the magnetic field.
The typical magnitude of the magnetic field in this region is $|B| \sim 0.7 B_*$
and the corresponding Larmor radius for the incoming upstream particles with $u=2.0$
is $r_g \sim 2.9$,
which is comparable to the size of the magnetic field structure,
i.e., the current filaments, in the transition region.
Therefore,
the incoming upstream particles should be isotropised in this region
because they are deflected at a large angle by the strong magnetic field at this location.
This isotropisation provides an effective dissipation mechanism for the upstream bulk kinetic energy.

\begin{figure}
\plotone{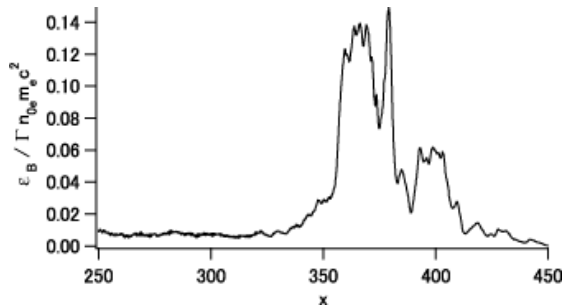}
\caption{ 
		\label{fig:UB}
		The $y$-averaged profile of the magnetic energy density
		normalized by the upstream bulk kinetic energy density at $t = 233$,
		$\epsilon_B / \Gamma n_{e0} m_e c^2$.
		The magnetic field strength reaches a sub-equipartition level
		at the peak and rapidly decays in the downstream region.
}
\end{figure}

Figure~\ref{fig:UB} shows the $y$-averaged profile
of the magnetic energy density normalized by the upstream
bulk kinetic energy density. We see that
the magnetic energy density
reaches about 10\% of the upstream bulk kinetic energy density
at the peak and decays downstream of the shock.
The saturation and decay mechanism of the magnetic field
would be similar to those of the Weibel instability \citep{Kato05}.
The current and the magnetic field generated by the instability in each filament
saturates when the magnetic field becomes strong enough to deflect
the current-carrying particles in the filament.
This predicts the sub-equipartition magnetic field
at saturation and it is consistent with the simulation result.
After the saturation,
the current filaments will coalesce with each other to
form larger filaments with the decreasing magnetic field strength.
We observe the evolution of the structure to a larger scale
together with the decay of the magnetic field strength
downstream of the shock transition region
in Figs.~\ref{fig:Bz_Jx}a, \ref{fig:profile}c, and \ref{fig:UB}.

\begin{figure}
\plotone{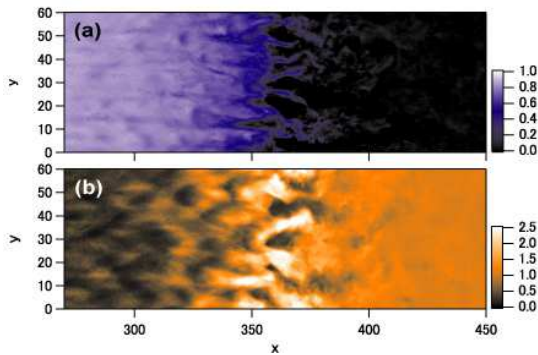}
\caption{ 
		\label{fig:Vx_KE}
		(a) The $x$ component of the electron mean velocity $V_x/c$ and
		(b) the mean kinetic energy of the electrons measured
		in the local plasma rest frame of the electrons,
		$\langle \gamma - 1 \rangle_\mathrm{fluid}$,
		at $t=233$.
		It is evident that the bulk kinetic energy of the upstream plasma
		is converted into the internal kinetic energy
		within the transition region.
		[\textit{See the electronic edition of the Journal for a
		color version of this figure}.]
}
\end{figure}

Figure~\ref{fig:Vx_KE} shows (a) the fluid velocity of electrons
in the $x$-direction, $V_x/c$,
and (b) the mean kinetic energy of the electrons measured
in the \textit{local} fluid rest frame of the electrons
expressed in terms of electron rest mass, $\langle\gamma - 1\rangle_\mathrm{fluid}$.
It is evident that
the incoming plasma from upstream
is rapidly decelerated within the transition region,
where the strong magnetic field exists
(see also Fig.~\ref{fig:Bz_Jx}a),
and its bulk kinetic energy is converted into
random or ``thermal'' kinetic energy in the downstream rest frame.

\begin{figure}
\plotone{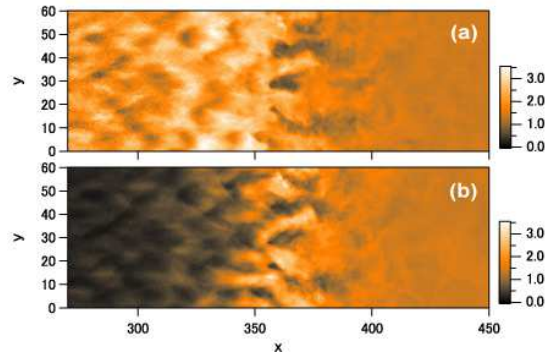}
\caption{ 
		\label{fig:sgmx_sgmy}
		Standard deviation of (a) $u_x$ and (b) $u_y$
		of electrons both measured in the local plasma rest frame
		of the electrons.
		The electrons are isotropised when they pass through the transition region.
		[\textit{See the electronic edition of the Journal for a
		color version of this figure}.]
}
\end{figure}

Figure~\ref{fig:sgmx_sgmy} shows
the standard deviations of (a) $u_x$ of electrons, $\sigma_x$,
and (b) $u_y$ of electrons, $\sigma_y$,
both measured in the local plasma rest frame
of the electrons.
We observe that the particles are isotropised
in the transition region due to
the strong magnetic field existing there.
It should be noted that,
in this two-dimensional simulation,
particles are isotropised only on the $x-y$ plane
and not in the $z$ direction
\citep[c.f.][]{Haruki03}
because only the Weibel modes with wave vectors
on the $x-y$ plane can develop
and hence only $B_z$ is generated.
In a three-dimensional simulation,
the Weibel modes with wave vectors
in the $z$ direction should develop equally
and particles are isotropised in all directions.

In a macroscopic view,
the features shown in the above figures
would be observed as those of shock waves;
the shock waves dissipate the upstream bulk kinetic energy
in the shock transition region
and propagate at an almost constant speed
into the upstream plasma.
In this sense, we can regard this as a collisionless shock.
In this shock,
the magnetic field in the transition region
generated by the Weibel-like instability
plays an essential role in the dissipation process
of the collisionless shock in unmagnetized plasmas.
It should be noted that
the shock has no corresponding fundamental linear wave,
which is required to define the Mach number,
in contrast to, for example, the perpendicular shocks in magnetized plasmas,
in which the magnetosonic wave is the fundamental
linear wave of the shock.
This point is clear from the fact that
the wave vector of the magnetic field in the transition region
is in the $y$ direction and not in the $x$ direction 
or the shock normal in the present case
and hence it has an essentially a two-dimensional structure.
Thus,
it can be said that this is a class of ``instability-driven''
shock waves.
For such shocks,
the concept of the Mach number
would not be applicable, and
the propagation speed of the shock
and the width of the transition region
would be essentially determined by the nature
of the instability that provides the dissipation mechanism
at the shock transition region.

\subsection{Charge-separated Current Filaments and Backward Flowing Particles}

\begin{figure}
\plotone{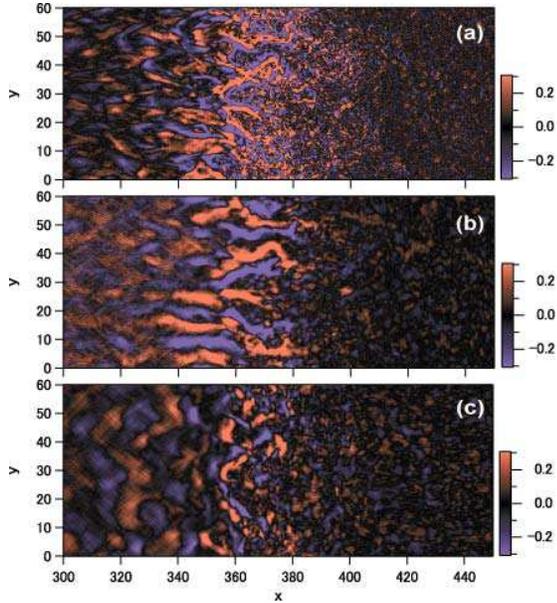}
\caption{ 
		\label{fig:rho_Ex_Ey}
		(a) Charge density $\rho$,
		(b) the $y$ component of the electric field $E_y$ and
		(c) the $x$ component of the electric field $E_x$ at $t=233$.
}
\end{figure}

From Fig.~\ref{fig:Bz_Jx}b,
we observe that there are current filaments even in the upstream region ($x < 350$).
Figure~\ref{fig:rho_Ex_Ey}a shows the charge density $\rho$.
Comparing these two figures,
it is evident that these filaments have a net space charge.
Since in each filament, $J_x$ and $\rho$ have the same sign,
these filaments are mainly composed of
the incoming upstream particles with $u_x > 0$.
Figure~\ref{fig:rho_Ex_Ey}b shows
the $y$ component of the electric field, $E_y$.
We observe that $E_y$ is generated between these filaments
due to their space charge.
As shown in Fig.~\ref{fig:Bz_Jx}a,
these filaments also generate magnetic field $B_z$ around them.
Thus,
rather strong electric and magnetic fields exist
in the upstream region, as shown in Fig.~\ref{fig:profile}c.
Figure~\ref{fig:rho_Ex_Ey}c shows
the $x$ component of the electric field.
We observed that strong electric fields exist
at the upstream edge of the transition region
and that there is a coherent structure in $E_x$
in the upstream of the transition region ($x < 350$).

\begin{figure}
\plotone{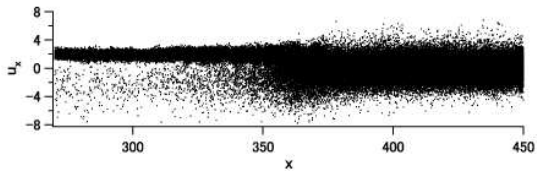}
\caption{ 
		\label{fig:electron_x_ux}
		The $x-u_x$ phase-space plot of electrons at $t=233$.
		Upstream of the shock transition region ($x < 350$),
		a fraction of particles flow backward, which causes
		two-stream type instabilities.
		In the downstream region ($x > 370$),
		electrons are thermalized (along the $x$ direction).
}
\end{figure}

The region where the charge-separated current filaments exist
would correspond to ``the charge separation layer''
suggested by \citet{Milosavljevic06}.
As they pointed out,
these filaments need a small fraction
of the particles flowing backward to stabilize themselves.
Figure~\ref{fig:electron_x_ux} shows
the $x-u_x$ phase-space plot of electrons at $t=233$.
We observe that
the particles flowing backward ($u_x < 0$) are supplied
around the shock transition region ($x \sim 350$).
These particles would stabilize
the charge-separated current filaments in the charge separation layer.
They can also drive the electrostatic beam instability.
The coherent electrostatic field $E_x$
upstream of the transition region
($x < 350$; see Fig.~\ref{fig:rho_Ex_Ey}c)
would be induced by this instability.
These particles should also contribute to drive the Weibel-type instability
around the transition region.

\begin{figure}
\plotone{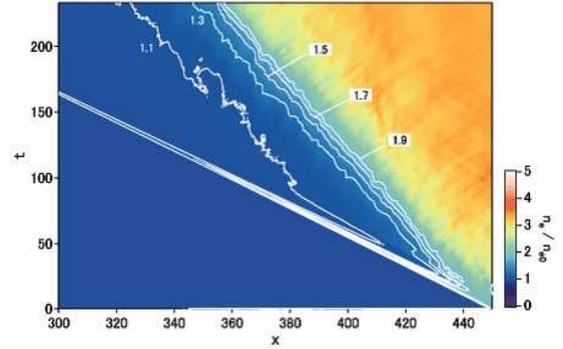}
\caption{ 
		\label{fig:backward_electron_x_t}
		Time development of the electron number density
		as in Fig.~\ref{fig:ne_history}, but with contours.
		The width of the precursor region ($1.1 < n_e/n_{e0} < 1.9$),
		where the density of the electrons flowing backward is high,
		is almost constant ($\Delta x \sim 30$) after $t \sim 100$.
		[\textit{See the electronic edition of the Journal for a
		color version of this figure}.]
}
\end{figure}

Figure~\ref{fig:backward_electron_x_t} shows
the time development of
the number density of electrons again
as in Fig.~\ref{fig:ne_history}, but with contours.
We observe that there is a ``precursor'' region in which the number density is
within a range of $1.1 < n_e/n_{e0} < 1.9$
in front of the transition region.
In this region,
the increase in the number density from that in the upstream plasma $n_{e0}$
would be mainly due to the electrons flowing backward
(see also Fig.~\ref{fig:electron_x_ux}).
The width of the precursor region
is almost constant ($\Delta x \sim 30$) after $t \sim 100$.

As already shown in Fig.~\ref{fig:Bz_Jx}b,
there are current filaments
in the transition region.
However,
they are charge-separated only at the filament edges
except those at the downstream side
that are continuously connected to the thermalized downstream region
(see Fig.~\ref{fig:rho_Ex_Ey}a).
The strength of the magnetic and electric fields,
current density, and charge density
all attain their maximum values in this region,
and the upstream kinetic energy
is dissipated mainly by the magnetic field
and partly by the electric field.

\begin{figure}
\plotone{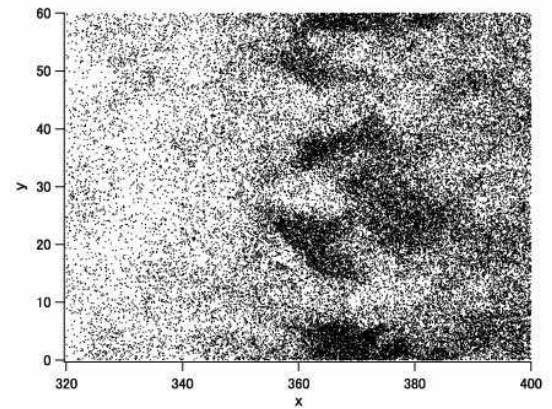}
\caption{ 
		\label{fig:electron_m2_x_y}
		Distribution of backward flowing electrons with $u_x < -2$.
		They move upstream through the current filaments with $J_x > 0$.
		This is evident for $x > 350$.
}
\end{figure}

A fraction of the downstream particles flowing backward with $u_x < u_\mathrm{sh,d} \sim -0.42$,
which is the 4-velocity corresponding to $V_\mathrm{sh,d}$,
can return upstream through the filaments
that carry the currents in the same direction as the particles.
On the other hand,
if the particles return through the filament carrying opposite currents,
the magnetic fields in the filaments
repel the particles from the filaments.
Figure~\ref{fig:electron_m2_x_y} shows the distribution of electrons
that are flowing upstream with $u_x < -2$.
By comparing with $J_x$ in Fig.~\ref{fig:Bz_Jx}b,
we observed that these electrons indeed flow
in the current filaments with $J_x > 0$.
In a similar manner,
the positrons flow backward in the current filaments,
but with $J_x < 0$ (not shown here).
However,
only a small fraction of the backward flowing
particles can move to the upstream region ($x < 350$)
because large potential barriers exist,
caused by the charge density
at the upstream edge of the filaments
(see also Figs.~\ref{fig:rho_Ex_Ey}a and \ref{fig:rho_Ex_Ey}c).
The particles that could pass through to the upstream region
would contribute to stabilize the charge-separated current filaments
in the charge separation layer, as mentioned earlier.

\subsection{Particle Energy Distribution and Acceleration}

\begin{figure}
\plotone{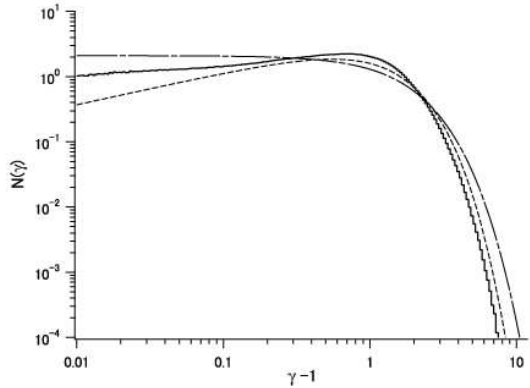}
\caption{ 
		\label{fig:hist_KE_downstream}
		Histogram of the kinetic energy of the downstream electrons
		within $430 < x < 442$ at $t=233$ (solid line).
		The horizontal axis is the kinetic energy normalized by
		the electron rest mass energy, $\gamma - 1$.
		The dashed curve and the dot-dashed curve
		represent the Maxwell distribution
		for three and two dimensions, respectively.
}
\end{figure}

Figure~\ref{fig:hist_KE_downstream}
shows the energy histogram of the electrons
accumulated over the area between $x=430$ and $x=442$
in the downstream region (the solid histogram).
The dashed curve and the dot-dashed curve
represent the Maxwell distributions
in three and two dimensions, which are defined by
\begin{equation}
	N_{3D}(\gamma) = n_d \frac{\theta^2 \gamma (\gamma^2-1)^{1/2}}{2 K_1(\theta)
								 + \theta K_0(\theta)} \exp(-\gamma \theta),
\end{equation}
and
\begin{equation}
	N_{2D}(\gamma) = n_d \frac{\theta^2 \gamma}{\theta + 1}
									 \exp\left[ -(\gamma-1) \theta\right],
\end{equation}
respectively.
Here, $n_d$ is the number density in the downstream region,
$K_i$ is the modified Bessel function of the second kind \citep{Abramowitz},
and $\theta$ is determined
so that $\int_1^\infty \gamma N(\gamma) d\gamma = \langle \gamma \rangle \sim \sqrt{5}$,
which is expected if the upstream bulk kinetic energy is completely converted
into the internal kinetic energy in the downstream rest frame.
From Fig.~\ref{fig:profile}a, we consider $n_d \sim 3.3\ n_{e0}$.
It is observed that the electron distribution
fits neither of the Maxwell distributions.

\begin{figure}
\plotone{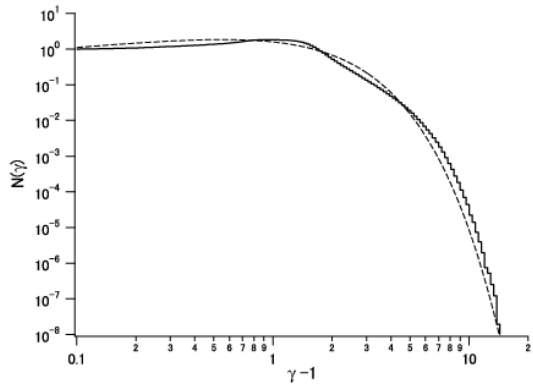}
\caption{\label{fig:hist_KE_all}
		Histogram of the kinetic energy of all the electrons
		in the simulation box (solid line).
		The three-dimensional Maxwell distribution
		is also plotted as a reference (dashed curve).
		There is no significant particle acceleration.
}
\end{figure}

Figure~\ref{fig:hist_KE_all} shows
the energy histogram of all electrons
in the simulation box by a solid histogram.
For reference,
we plotted the three-dimensional Maxwell distribution
by the dotted curve.
Their normalizations are arbitrary,
and there is no significant particle acceleration.

\section{Conclusions}
The simulation results clearly show
that collisionless shocks can be driven
in unmagnetized electron-positron plasmas.
The structure of the collisionless shock
propagates at an almost constant speed.
The dissipation of the upstream bulk kinetic energy
is mainly due to the deflection of particles
by the strong magnetic field
generated around the shock transition region
by the Weibel-type instability.
It is remarkable that
this shock has no corresponding linear wave and therefore
it can be regarded as a kind of ``instability-driven'' shock wave.
We also observed the electric and magnetic field generation
even in the upstream of the shock transition region
due to the charge-separated current filaments
together with the electrostatic beam instability.
We found that a fraction of the downstream thermalized particles
return upstream through the shock transition region.
These particles would interact with the upstream incoming plasma
and cause
the generation of the charge-separated current filaments
as well as
the electrostatic beam instability
upstream of the transition region.
We did not observe any efficient particle acceleration processes in our simulation,
whereas we observed relatively rapid decay of the generated magnetic field
within the shock transition region.
On the other hand, in the real world, there are many evidences to suggest that
particles are accelerated at shocks, and there are several indications that large-scale,
sub-equipartition magnetic fields may be generated around these shocks.
The reasons why these processes were not observed in our simulation may be that
(1) the background magnetic field is essential for them,
(2) the spatial and temporal scales of the simulation are too small to deal with them,
or
(3) these processes are inefficient in the electron-positron shocks.

\acknowledgments
I would like to thank F. Takahara, M. Hattori, Y. Fujita, and S. Nagataki
for their helpful discussions.
This research was supported in part by
the Japan Science and Technology Agency.
Numerical computations were carried out on VPP5000 at the Astronomical Data Analysis Center, ADAC,
of the National Astronomical Observatory of Japan.


\end{document}